\documentclass{PoS}
\pdfoutput=1
\usepackage[UKenglish]{babel}
\usepackage{natbib}
\newcommand{\gev}{\, {\rm GeV}}
\newcommand{\mev}{\, {\rm MeV}}
\newcommand{\be}{\begin{equation}}
\newcommand{\ee}{\end{equation}}
\newcommand{\bea}{\begin{eqnarray}}
\newcommand{\eea}{\end{eqnarray}}

\title{A determination of the average up-down, strange and charm quark masses from $N_f=2+1+1$}

\ShortTitle{A determination of the average up-down, strange and charm quark masses from $N_f=2+1+1$}

\author{ N. Carrasco$^{(a)}$, P. Dimopoulos$^{(b,c)}$, R. Frezzotti$^{(c,d)}$, \speaker{P. Lami}$^{(e,f)}$, V. Lubicz$^{(e,f)}$, 

D. Palao$^{(g)}$, E. Picca$^{(e,f)}$, L. Riggio$^{(e,f)}$, G.C. Rossi$^{(c,d)}$, F. Sanfilippo$^{(h)}$, S. Simula$^{(f)}$,
       
C. Tarantino$^{(e,f)}$
\\

\it $^{(a)}$ Departament de F\'isica Te\`orica and IFIC, Univ. de Val\`encia-CSIC, Av. Dr. Moliner 50, E-46100
Val\`encia, Spain, E-mail: \email{nuria.carrasco@uv.es}

\it $^{(b)}$ Centro Fermi - Museo Storico della Fisica e Centro Studi e Ricerche E. Fermi, Rome, Italy

\it $^{(c)}$ Dipartimento di Fisica, Universit\`a di Roma ``Tor Vergata'', Rome, Italy.     E-mail:\email{dimopoulos@roma2.infn.it}, \email{frezzotti@roma2.infn.it}, \email{rossig@roma2.infn.it} 

\it $^{(d)}$ INFN, Sezione di ``Tor Vergata", Rome, Italy

\it $^{(e)}$ Dipartimento di Matematica e Fisica, Universit\`a  Roma Tre, Rome, Italy.             

Email: \email{lamipaolo@gmail.com}, \email{lubicz@fis.uniroma3.it}, \email{e.picca88@gmail.com}, \email{lorenzo.riggio@gmail.com}, \email{tarantino@fis.uniroma3.it}

\it $^{(f)}$ INFN, Sezione di Roma Tre, Rome, Italy. Email: \email{simula@roma3.infn.it}

\it $^{(g)}$ University of Frankfurt, Frankfurt, Germany. 

Email: \email{palao@th.physik.uni-frankfurt.de}

\it $^{(h)}$ Laboratoire de Physique Th\'eorique (B$\hat{a}$t. 210),
\it Universit\'e Paris Sud, F-91405 Orsay-Cedex, France. Email: \email{fr.sanfilippo@gmail.com}

\\

\bf{For the ETM Collaboration}
}

\abstract{We present a lattice QCD determination of the average up-down, strange and charm quark masses based on simulations performed by the European Twisted Mass Collaboration with $N_f = 2 + 1 + 1$ dynamical fermions. We simulated at three different values of the lattice spacing,
the smallest being approximately $0.06fm$, and with pion masses as small as $210 \mev$. Our results are: $m_{ud}(2\gev)=3.70(17)\mev$, $m_s(2\gev)=99.2(3.9)\mev$, $m_c(m_c)=1.350(49)\gev$, $m_s/m_{ud}=26.64(30)$ and $m_c/m_s=11.65(12)$.}

\FullConference{31st International Symposium on Lattice Field Theory LATTICE 2013\\
                 July 29 - August 3, 2013\\
                 Mainz, Germany}

\begin{document}

\section{Introduction and simulation details}
\label{sec:intro}

The precise knowledge of the quark masses and in general of the hadronic parameters plays a fundamental role in testing the Standard Model (SM) and in the search for new physics. 

In this contribution we present an accurate determination of the average up-down, strange and charm quark masses using the gauge configurations produced by the European Twisted Mass (ETM) Collaboration with four flavors of dynamical quarks ($N_f = 2+1+1$), which include in the sea, besides two light mass-degenerate quarks, also the strange and the charm quarks, with masses close to their physical values \cite{Baron:2010bv,Baron:2011sf}. 
The simulations were carried out at three different values of the inverse bare lattice coupling $\beta$ which allow for a controlled extrapolation to the continuum limit. 
For each ensemble we used a subset of well-separated trajectories to avoid autocorrelations. 
The Iwasaki action \cite{Iwasaki} was adopted for the gauge links, while the sea fermions were simulated using the Wilson Twisted Mass Action \cite{Frezzotti:2003xj}, which at maximal twist provides automatic ${\cal{O}}(a)$-improvement \cite{Frezzotti:2003ni}. 
To avoid mixing in the strange and charm sectors we adopted a non-unitary setup in which for each flavor valence quarks are simulated using the Osterwalder-Seiler action \cite{Osterwalder:1977pc}.
In order to minimize discretization effects in the pseudoscalar (PS) meson masses the values of the Wilson parameter $r$ are always chosen so that the two valence quarks in a meson have opposite values of $r$.

At each lattice spacing different values of the light sea quark masses have been considered. 
The light valence and sea quark masses are always degenerate. 
In the light, strange and charm sectors the quark masses were simulated in the ranges $0.1 ~ m_s^{phys} \lesssim  \mu_l \lesssim 0.5 ~ m_s^{phys}$, $0.7 ~ m_s^{phys} \lesssim  \mu_s \lesssim 1.2 ~ m_s^{phys}$ and $0.7 ~ m_c^{phys} \lesssim  \mu_c \lesssim 2.0 ~ m_c^{phys}$, respectively.

We studied the dependence of the PS meson masses (and of the pion decay constant) on the (renormalized) light quark mass fitting simultaneously the data at different lattice spacings and volumes. 
The values of the lattice spacing obtained in our analysis are $a = 0.0885(36)$, $0.0815(30)$, $0.0619(18)$ fm, and the lattice volume goes from $\simeq 2$ to $\simeq 3$ fm. 
The pion masses, extrapolated to the continuum and infinite volume limits, range from $\simeq 210$ to $ \simeq 450 \mev$.

Within our analyses we used: ~ i) the results for $r_0/a$ ($r_0$ is the Sommer parameter \cite{Sommer:1993ce}) obtained extrapolating to the chiral limit the values computed in \cite{Baron:2010bv,Baron:2011sf}, and ~ ii) the quark mass renormalization constant $Z_m = 1/Z_P$ computed non-perturbatively in the RI-MOM scheme.

\section{Pion analysis}
We determined the value of the average up-down quark mass $m_{ud}$ by studying the dependence of the squared PS meson mass $M_{\ell\ell}^2$ on the (renormalized) light quark mass $m_\ell$ and on the lattice spacing $a$.
The pion decay constant $f_{\pi}$ was used to set the scale.

In the case of twisted mass fermions at fixed lattice spacing and volume the ChPT predicts for the quantities $M_{\ell\ell}^2$ and $f_{\ell\ell}$ the following behavior at the next-to-leading order (NLO)
 \bea
     \label{eq:cptmpi2Ch}
     M_{\ell\ell}^2  & = & 2 B_0 m_\ell \left( {1 + \xi _l \log \xi _l  + P_1 \xi _l  + P_2 ~ a^2 + \frac{{4c_2 a^2}}{{(4\pi f_0)^2 }} \log \xi _l } \right) 
                                     \cdot K_M^{FSE} \quad \\
    \label{eq:cptfpiCh}
    f_{\ell\ell} & = & f_0 \left( {1 - 2\xi _l \log \xi _l  + P_3 \xi _l  + P_4 ~ a^2 - \frac{{4c_2 a^2 }}{{(4\pi f_0)^2 }} \log \xi _l } \right) \cdot K_f^{FSE} , ~
 \eea
where $\xi_l = 2B_0m_\ell / (4\pi f_0)^2$ with $B_0$ and $f_0$ being the LO low-energy constants (LECs).
In Eqs.~(\ref{eq:cptmpi2Ch})-(\ref{eq:cptfpiCh}) the terms proportional to $a^2 \log(\xi_l)$ originate from the mass splitting between charged and  neutral pions, which can be expressed in terms of the $c_2$ parameter as $(M^2 _{\pi ^0 }  - M^2 _{\pi ^ \pm  } )_{LO} = 4a^2 c_2$.
They represent an additional discretization effect stemming from the twisted-mass fermionic action \cite{Bar:2010jk,Herdoiza:2013sla}.
The factors $K_M^{FSE}$ and $K_f^{FSE}$ represent the corrections for the finite size effects (FSE) to $M_{\ell\ell}^2$ and $f_{\ell\ell}$, respectively, computed by Colangelo, Wenger and Wu (CWW) \cite{Colangelo:2010cu}.
\begin{figure}[t!]
\centering
\scalebox{0.25}{\includegraphics{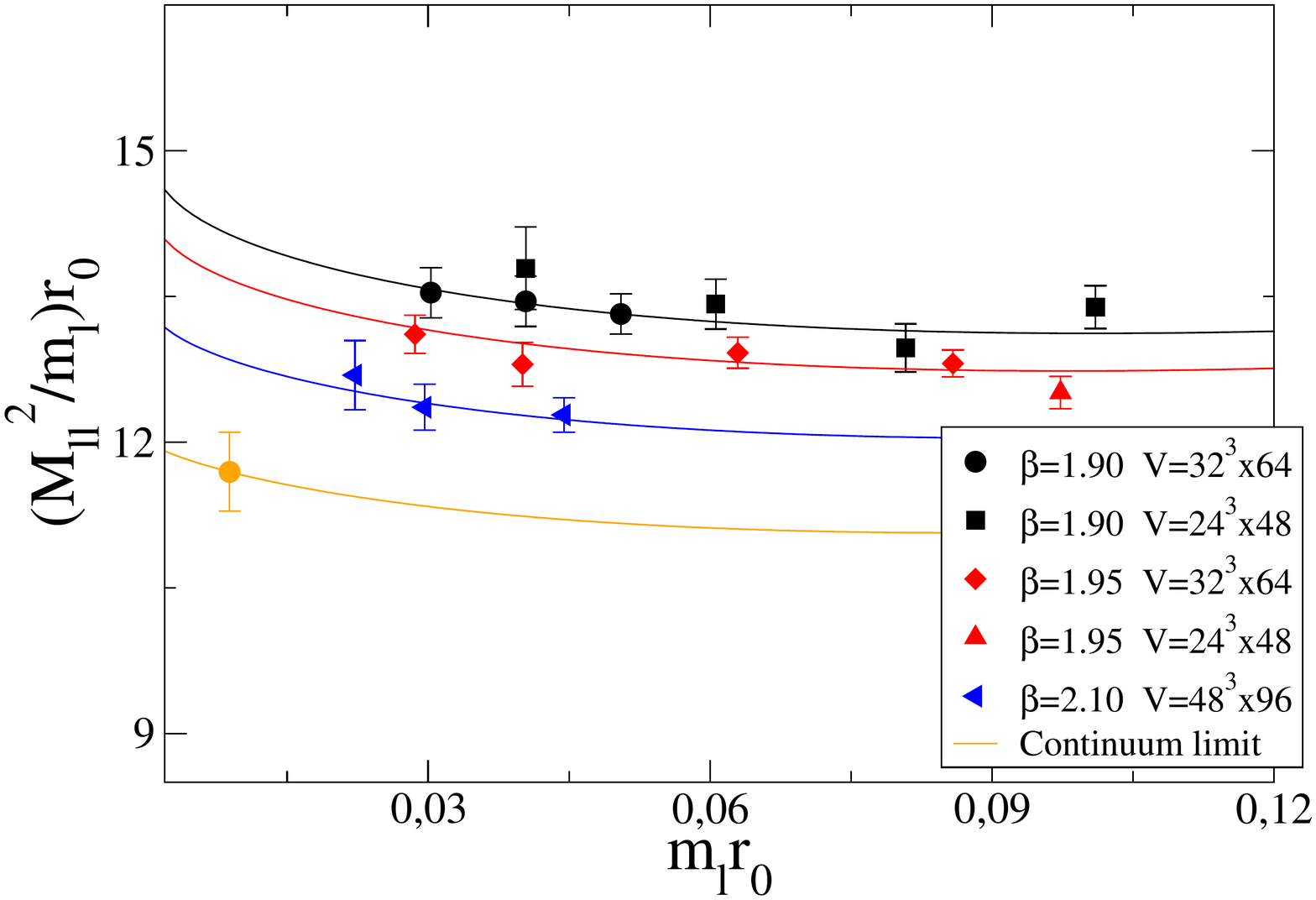} \includegraphics{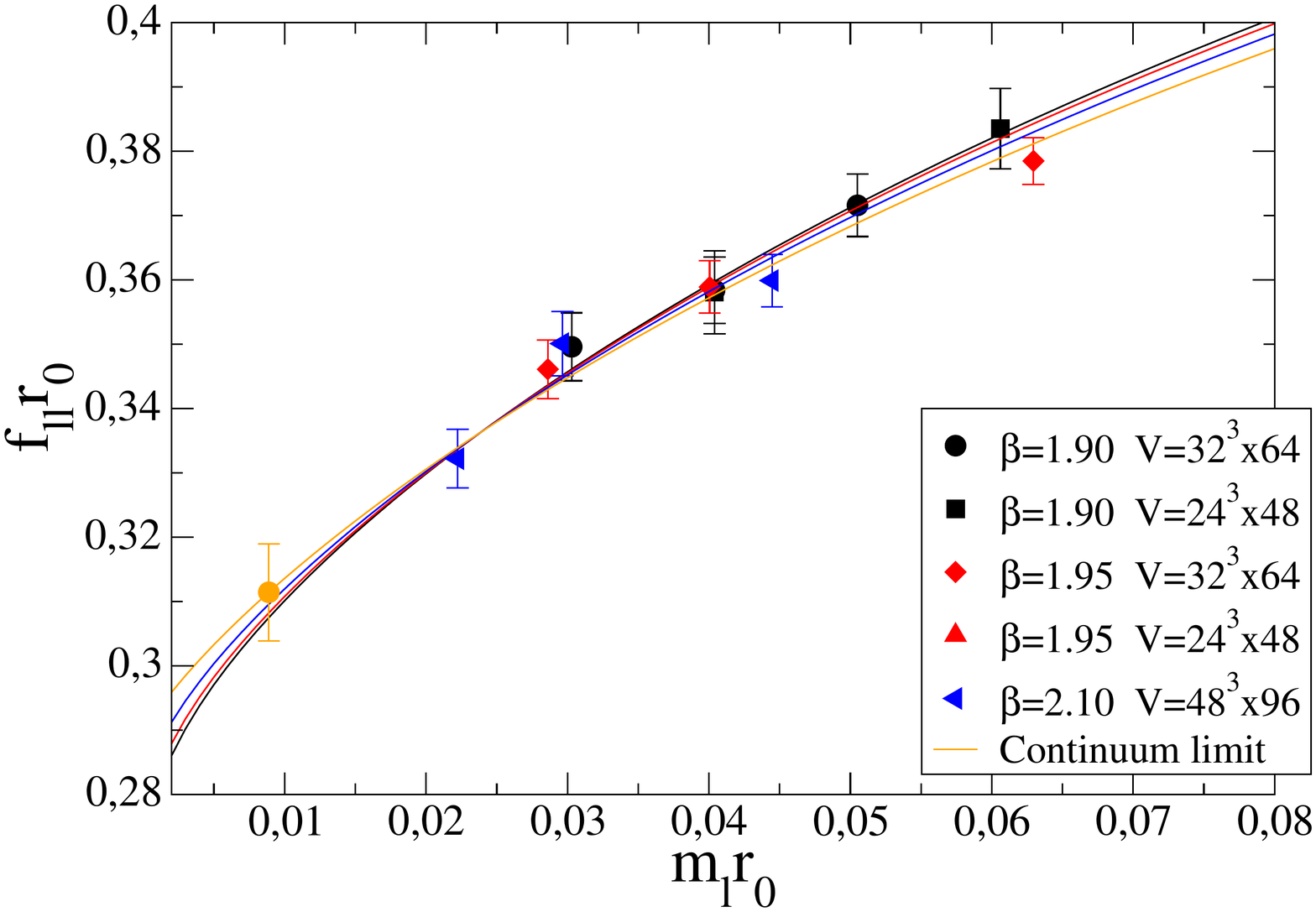}}
\caption{\it Chiral and continuum extrapolations of $r_0 M_{\ell\ell}^2 / m_\ell$ (left) and of $r_0 f_{\ell\ell}$ (right) based on the NLO ChPT fits given in Eqs.~(\ref{eq:cptmpi2Ch})-(\ref{eq:cptfpiCh}). Lattice data are corrected for FSE using the CWW approach \cite{Colangelo:2010cu}.}
\label{fig:mpi2sumlmlCh}
\end{figure}

The chiral and continuum extrapolations of $r_0 M_{\ell\ell}^2 / m_\ell$ and of $r_0 f_{\ell\ell}$ are shown in Fig.~\ref{fig:mpi2sumlmlCh}. It can be seen that the impact of discretization effects when we use $r_0$ as the scaling variable is at the level of $\simeq 10 \%$ for $r_0 M_{\ell\ell}^2 / m_\ell$.
In order to keep the extrapolation to the continuum limit under better control we performed an alternative analysis adopting a different choice for the scaling variable, namely instead of $r_0$ we considered the mass $M_{s^\prime s^\prime}$ of a fictitious PS meson composed by two strange-like valence quarks with mass fixed at $r_0 m_{s^\prime} = 0.22$. 
The mass $M_{s^\prime s^\prime}$ is affected by non-negligible cutoff effects, similar to the ones of the K-meson without however any significant dependence on the light quark mass. 
Thus, we performed the continuum extrapolation of the ratio $M_{\ell\ell}^2 / M_{s^\prime s^\prime}^2$, which benefits of a (partial) cancellation of discretization effects.
The comparison between the analyses performed in units of $r_0$ and $M_{s^\prime s^\prime}$ shows that, when $M_{s^\prime s^\prime}$ is chosen as the scaling variable, the discretization effects on $M_{\ell\ell}^2$ can be reduced from $\simeq 10 \%$ down to $\simeq 4.5 \%$.

\begin{figure}[t!]
\centering
\scalebox{0.30}{\includegraphics{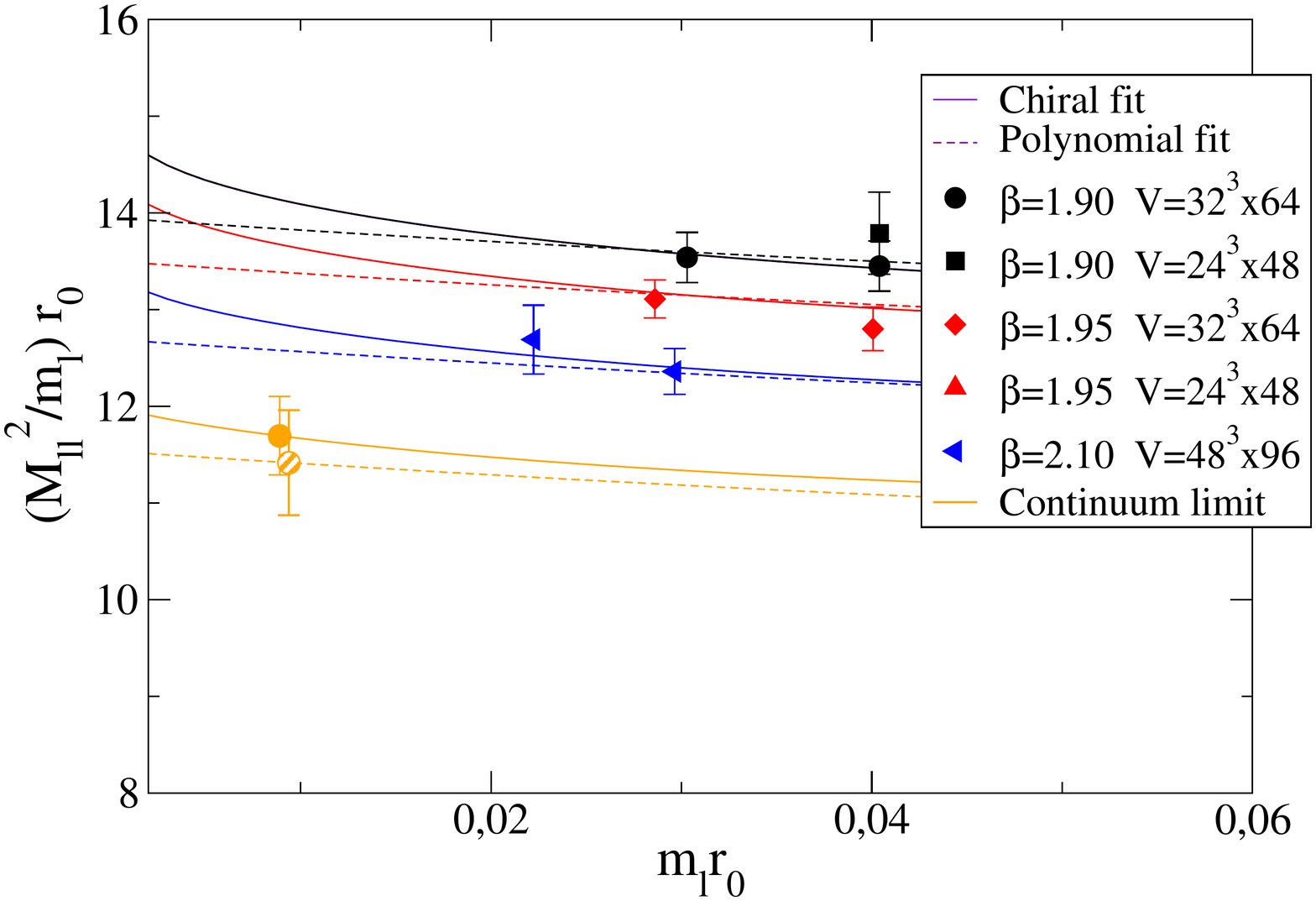}}
\caption{\it Comparison between NLO ChPT and polynomial fits for $r_0 M_{\ell\ell}^2 / m_\ell$ as a function of $m_\ell$.}
\label{fig:mpi2comp}
\end{figure}

For the chiral extrapolation we adopted both the NLO ChPT predictions (\ref{eq:cptmpi2Ch})-(\ref{eq:cptfpiCh}) and a polynomial formula. 
The corresponding results are presented in Fig.~\ref{fig:mpi2comp}.
The various sources of systematic uncertainties are estimated as follows.
The difference of the results obtained using either $r_0$ or $M_{s^\prime s^\prime}$ as the scaling variable represent the systematic uncertainty on $m_{ud}$ associated with discretization effects.
A systematic uncertainty related to the chiral extrapolation is obtained by comparing the results  of the NLO ChPT predictions with those of the polynomial ansatz.
As for FSE we compared the results obtained by applying the CWW corrections with the ones obtained without correcting for FSE. 
Finally, the methods M1 and M2 (differing by ${\cal{O}}(a^2)$ effects), used in \cite{ETM:2011aa} to calculate the renormalization constant $Z_P$ in the RI-MOM scheme, allow us to estimate the systematic uncertainty due to the mass renormalization constant.

Combining all sources of uncertainties we get the following estimate for the average up-down quark mass in the ${\overline{\rm MS}}$ scheme at a renormalization scale of $2 \gev$:
 \bea
    m_{ud}   =  3.70(13)_{stat + fit} (6)_{Chiral} (5)_{Disc.} (5)_{Z_P} (4)_{FSE} \mev  =  3.70(17) \mev  ~ .
    \label{eq:mudresults}
 \eea
Our determination for $m_{ud}$ is the first one obtained with $N_f = 2+1+1$. The recent lattice averages provided by FLAG \cite{FLAG} are: $m_{ud} = 3.6(2) \mev$ at $N_f = 2$ and $m_{ud} = 3.42(9) \mev$ at $N_f = 2+1$.

\section{Kaon analysis}
The physical value of the strange quark mass $m_s$ has been determined by studying the squared kaon mass $M_{s\ell}^2$ as a function of $m_s$, $m_\ell$ and $a^2$. 
The analysis followed a strategy similar to the pion one, adopting for FSE the formulae provided by Colangelo, Durr and Haefeli (CDH) \cite{CDH05}. 
As a first step, however, we performed an interpolation of the lattice kaon data at a fixed value of the strange quark mass in order to arrive iteratively to the physical value $m_s$. The latter is the one that leads to the value of the kaon mass in pure QCD, i.e.~the experimental one corrected for leading strong and electromagnetic isospin breaking effects, namely $\hat{M}_K = 494.2 \mev$ \cite{deDivitiis:2013xla,FLAG}. 

To determine the strange quark mass we used several quantities extracted from the pion sector, namely the lattice spacing, the LECs $B_0$ and $f_0$, the Sommer parameter $r_0$ and the result for the average up-down quark mass $m_{ud}$.
As in the case of the pion, systematic errors were estimated by comparing the results obtained with different procedures.

For the chiral extrapolation we used either a quadratic polynomial formula or the SU(2) ChPT prediction at NLO.
The latter one reads as
 \be
    M_{s\ell}^2  = P_0 (m_\ell  + m_s) \left[1 + P_1 m_\ell   + P_3 a^2 \right] \cdot K_{M_{s\ell}}^{FSE} ~ . 
    \label{eq:mk2Ch}
 \ee
due to the absence of chiral logs at NLO.
Therefore, the polynomial formula differs from Eq.~(\ref{eq:mk2Ch}) only for a quadratic term in $m_\ell$ inside the square brackets.

As in the pion case, the difference between the results of the two chiral extrapolations is taken as an estimate of the systematic uncertainty due to the chiral extrapolation, while the result for $m_s$ obtained including the CDH FSE corrections is compared with the one evaluated without FSE corrections to estimate the corresponding systematic error.

Following the same strategy adopted in the pion analyses the kaon masses simulated at different $\beta$ values can be brought to a common scale either by using $r_0$ or by constructing the ratios $M_{s\ell}^2 / M_{s^\prime s^\prime}^2$, which are expected to suffer only marginally by discretization effects.
In this case $m_\ell$ is expressed in physical units by using the values of the lattice spacing found in the pion sector.

\begin{figure}[t!]
\centering
\scalebox{0.25}{\includegraphics{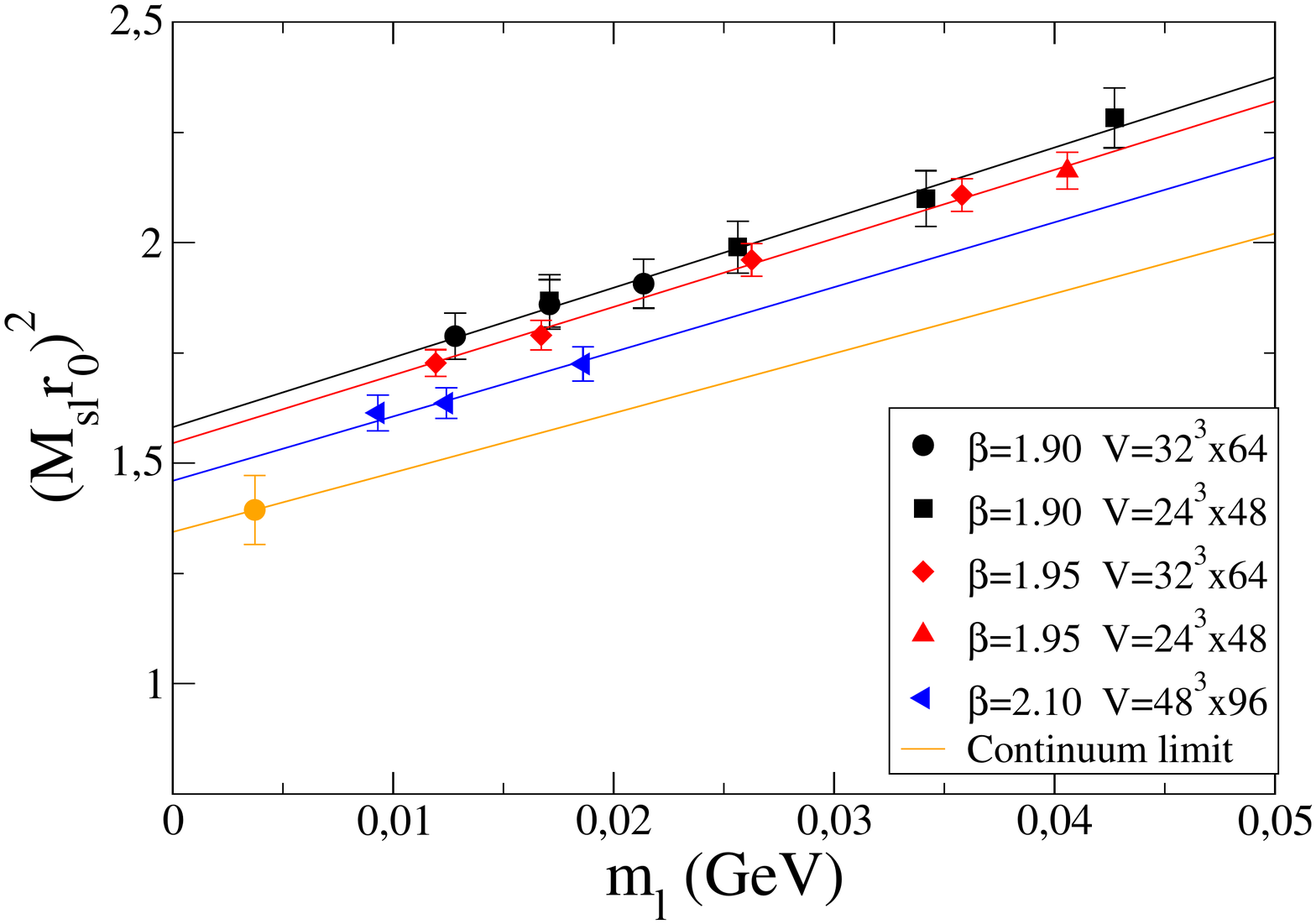} \includegraphics{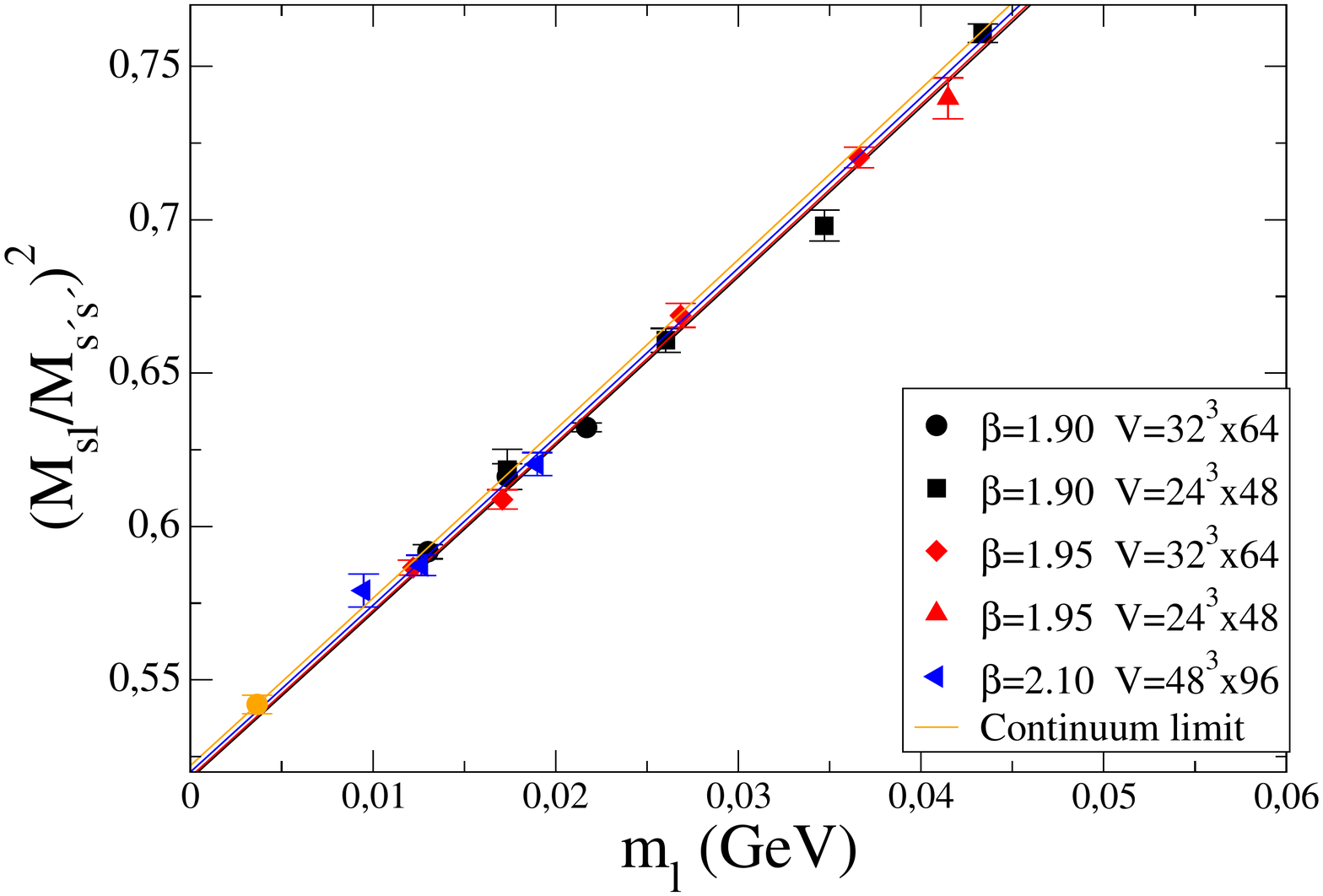}}
\caption{\it Chiral and continuum extrapolation of $(r_0 M_{s\ell})^2$ (left) and $(M_{s\ell} / M_{s^\prime s^\prime})^2$ (right) based on the NLO ChPT fit given in Eq.~(\ref{eq:mk2Ch}). Lattice data have been corrected for FSE using the CDH approach \cite{CDH05}.}
\label{fig:mk2Ch}
\end{figure}

The dependencies of $M_{s\ell}^2 r_0^2$ and $M_{s\ell}^2 / M_{s^\prime s^\prime}^2$ on the renormalized light quark mass at each values of $\beta$ as well as in the continuum limit are shown in Fig.~\ref{fig:mk2Ch} in the case of the $SU(2)$ ChPT fit.
Results of the same quality are obtained adopting the polynomial fit in both $r_0$ and $M_{s^\prime s^\prime}$ units.
In the case of the kaon mass the use of the hadronic scale $M_{s^\prime s^\prime}$ turns out to be an extremely efficient tool for an almost total cancellation of discretization effects in $M_{s\ell}^2$, namely from $\simeq 10\%$  to about $0.4 \%$ (see Fig.~\ref{fig:mk2Ch}).
The comparison between the two analyses based on $r_0$ or $M_{s^\prime s^\prime}$ as scaling variable allows us to estimate the systematic error due to discretization effects.

Our final result for $m_s$ in the ${\overline{\rm MS}}$ scheme at a renormalization scale of $2 \gev$ is
 \bea
    m_s  =  99.2 (3.4)_{stat + fit} (0.6)_{Chiral} (1.1)_{Disc.} (1.5)_{Z_P} (0.5)_{FSE} \mev  =  99.2 (3.9) \mev ~ .
    \label{eq:msresults}
 \eea

\section{$D_s$ analysis} 
The physical value of the charm quark mass $m_c$ has been extracted from the analysis of the $D_s$-meson mass as a function of $m_c$, $m_s$, $m_\ell$ and $a^2$.  
Again, as a first step we performed an interpolation of lattice data on $M_{hs}$ both at the physical strange quark mass (obtained from the kaon analysis) and at a fixed value of the charm quark mass in order to reach iteratively the physical value $m_c$, which is the one leading to the experimental value $M_{D_s}^{exp} = 1.969 \gev$.

As in the cases of the pion and kaon analyses, the lattice data for $M_{hs}$ are converted in units of either $r_0$ or the mass $M_{c^\prime s^\prime}$ of a fictitious PS meson, made with one valence strange-like and one valence charm-like quarks with masses fixed at $r_0 m_{s^\prime} = 0.22$ and $r_0 m_{c^\prime} = 2.4$, respectively. 
By construction such a reference mass $M_{c^\prime s^\prime}$ has discretization effects closer to the ones of $M_{hs}$.

\begin{figure}[t!]
\centering
\scalebox{0.25}{\includegraphics{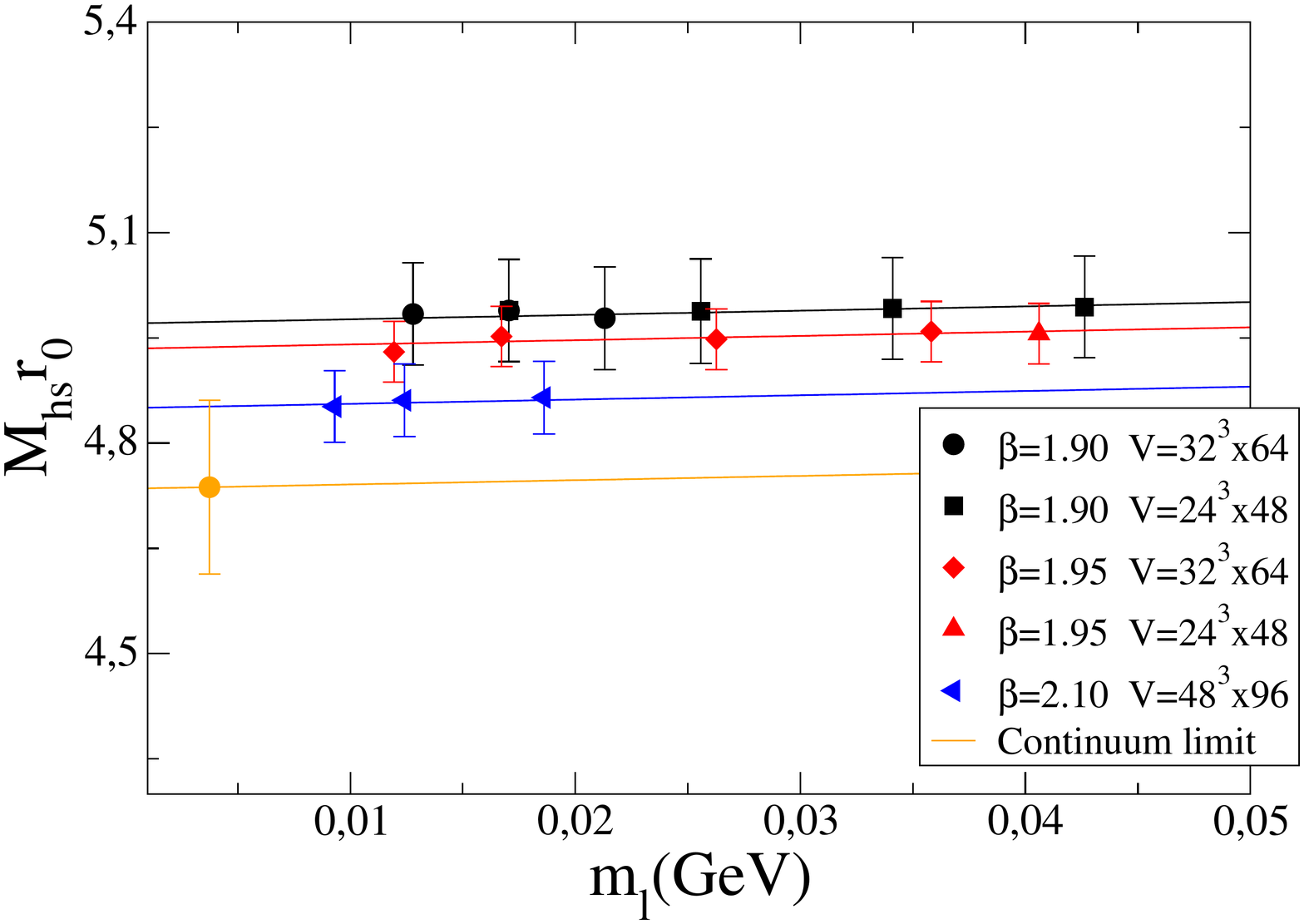} \includegraphics{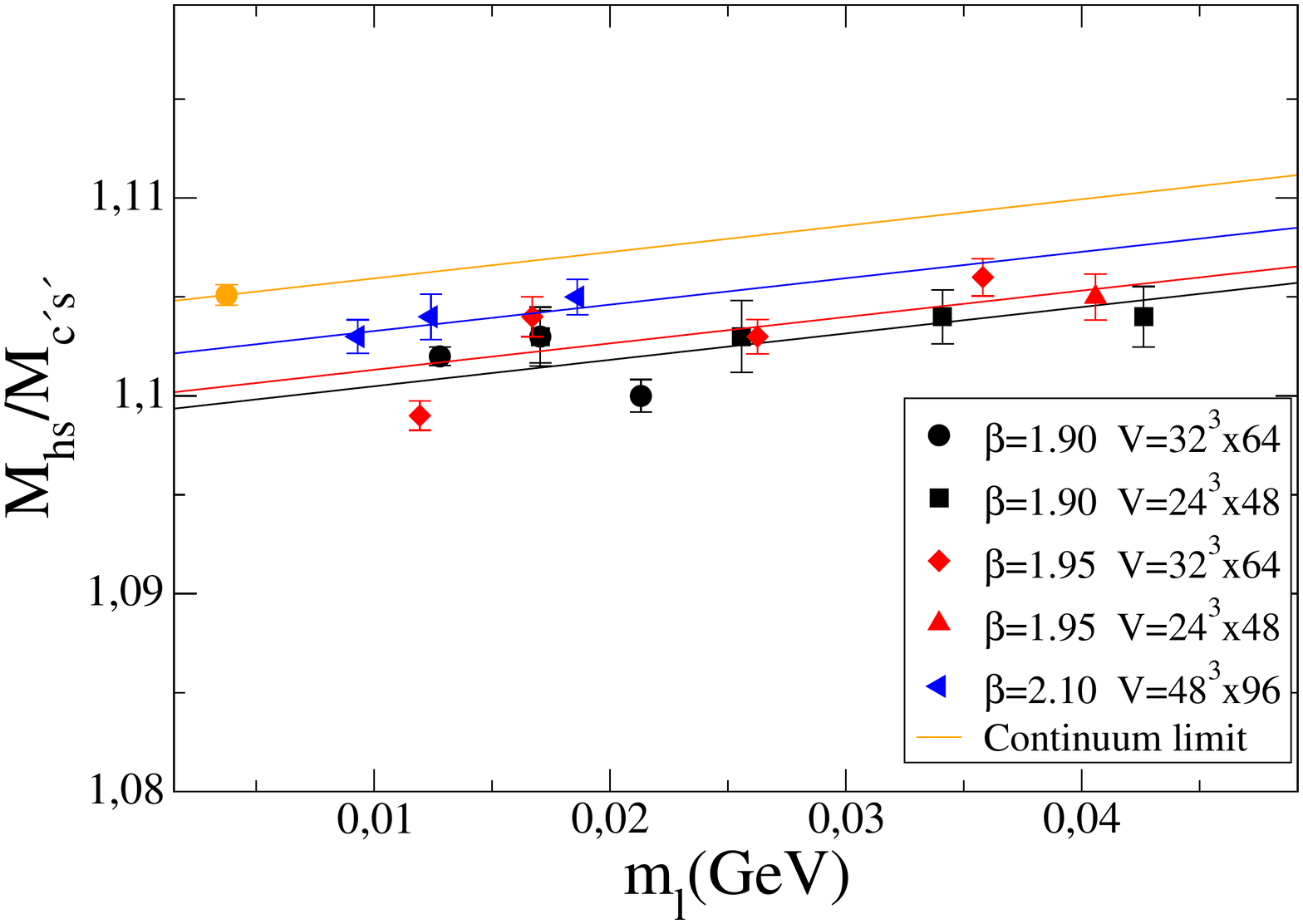}}
\caption{\it Chiral and continuum extrapolations of $r_0 M_{hs}$ (left) and $M_{hs}/M_{c^\prime s^\prime}$ (right) using a linear fit in $m_\ell$.}
\label{fig:mdslin}
\end{figure}

The chiral and continuum extrapolations of $r_0 M_{hs}$ and $M_{hs}/M_{c^\prime s^\prime}$ are shown in Fig.~\ref{fig:mdslin}.
The systematic uncertainty associated with the chiral extrapolation has been investigated using either a linear or a quadratic fit in $m_\ell$.
We have taken into account also the errors induced both by the uncertainty on $m_s$ and by the different $Z_P$ determinations from the methods M1 and M2.

After evolving the renormalization scale from $2 \gev$ to $m_c$ using $N^3LO$ perturbation theory with $N_f = 4$, we obtained
 \bea
      m_c (m_c )  = 1.350 (44)_{stat + fit + scale} (3)_{Chiral} (8)_{Disc.} (19)_{Z_p } (5)_{m_s } {\rm{GeV}}  =  1.350 (49) \gev ~ ,
     \label{eq:mcresults}
 \eea
where it can be seen that the largest uncertainty comes from the combination of the statistical error, of the fitting uncertainty and of the error on the lattice spacing found in the pion analyses.
 
\section{Determination of the ratios $m_s / m_{ud}$ and $m_c / m_s$}
In order to calculate the ratios $m_s/m_{ud}$ and $m_c/m_s$ one could simply use our previous results and propagate the uncertainties.
We considered however an alternative procedure in which various sources of uncertainties are minimized and a more precise determination of the mass ratios (at the percent level) can be obtained.
For determining the ratio $m_s/m_{ud}$ we have constructed the quantity 
 \be
    R(m_s, m_\ell, a^2) \equiv \frac{m_\ell}{m_s}  \frac{2 M_{s\ell}^2 - M_{\ell\ell}^2}{M_{\ell\ell}^2} ~ .
    \label{eq:R}
 \ee
After applying the FSE corrections to $M_{\ell\ell}^2$ and $M_{s \ell}^2$ described in the previous sections, the continuum and chiral extrapolations of $R$ provide its physical value $R^{phys} = R(m_s, m_{ud}, 0)$, in terms of which the ratio $m_s/m_{ud}$ can be calculated as 
 \be
     \frac{m_s}{m_{ud}} = \left( \frac{2 M_K^2 - M_\pi^2}{M_\pi^2} \right)^{phys} \frac{1}{R^{phys}} ~ ,
     \label{eq:ratiosl}
  \ee
Our result is
 \be
    \frac{m_s}{m_{ud}} = 26.64 (30)_{stat+fit} (2)_{a+m_\ell} (2)_{Z_P} (1)_{FSE} =26.64(30),
    \label{eq:msml}
 \ee
where all the systematic errors were calculated as in the previous analyses. The subscript $(a+m_\ell)$ indicates the systematic error induced by the uncertainties on both the lattice spacing and the light quark mass. The uncertainty induced by the error on $m_s$ turned out to be negligible.

A similar strategy has been implemented for the ratio $m_c/m_s$ using the quantity
 \be
    \overline{R}(m_c, m_s, m_\ell, a^2) \equiv \frac{m_s}{m_c}  \frac{(M_{cc}-M_{cs})(2M_{cs}-M_{cc})}{2 M_{s \ell}^2 - M_{\ell\ell}^2} ~ .
    \label{eq:Rbar}
 \ee
We got the preliminary result
 \be
      \frac{m_c}{m_s} =11.65 (9)_{stat+fit} (5)_{m_s} (6)_{m_c} = 11.65 (12) ~ ,
      \label{eq:mcms}
 \ee
where the systematic error associated to the uncertainties on the lattice spacing and the light quark mass has not yet been included. 

\section*{Acknowledgements}
We acknowledge the CPU time provided by the PRACE Research Infrastructure under the project PRA027 ``QCD Simulations for Flavor Physics in the Standard Model and Beyond'' on the JUGENE BG/P system at the J\"ulich SuperComputing Center (Germany), and by the agreement between INFN and CINECA under the specific initiative INFN-RM123.

D.P. wants to thank the HIC for FAIR within the framework of the LOEWE program launched by the State of Hesse for partial support.

\end{document}